\documentclass{article}

\usepackage[dvips]{graphicx}
\usepackage{amssymb,amsfonts,amsmath}
\begin{document}

\title{Fundamental Limits to Position Determination by Concentration
Gradients}

\author{Filipe Tostevin$^1$, Pieter Rein ten Wolde$^2$ and Martin
Howard$^{1*}$\\
\\
$^1$ Department of Mathematics, Imperial College London, \\
South Kensington Campus, London SW7 2AZ, UK \\
$^2$ FOM Institute for Atomic and Molecular Physics (AMOLF), \\
Kruislaan 407, 1098 SJ, Amsterdam, The Netherlands \\
\\
$^*$ To whom correspondence should be addressed.\\
E-mail:martin.howard@imperial.ac.uk
}

\maketitle

\begin{abstract}
Position determination in biological systems is often achieved through
protein concentration gradients. Measuring the local concentration of
such a protein with a spatially-varying distribution allows the
measurement of position within the system. In order for these systems
to work effectively, position determination must be robust to
noise. Here, we calculate fundamental limits to the precision of
position determination by concentration gradients due to unavoidable
biochemical noise perturbing the gradients. We focus on gradient
proteins with first order reaction kinetics. Systems of this type have
been experimentally characterised in both developmental and cell
biology settings. For a single gradient we show that, through
time-averaging, great precision can potentially be achieved even with
very low protein copy numbers. As a second example, we investigate the
ability of a system with oppositely directed gradients to find its
centre. With this mechanism, positional precision close to the centre
improves more slowly with increasing averaging time, and so longer
averaging times or higher copy numbers are required for high
precision. For both single and double gradients, we demonstrate the
existence of optimal length scales for the gradients, where precision
is maximized, as well as analyzing how precision depends on the size
of the concentration measuring apparatus.  Our results provide
fundamental constraints on the positional precision supplied by
concentration gradients in various contexts, including both in
developmental biology and also within a single cell.
\end{abstract}

\section*{Summary}

In many biological systems gradients of protein concentration provide
precise positional information.  Above a critical concentration a
signal can be switched on by the gradient, whereas below the threshold
the signal is switched off, thereby providing position dependent
signalling. Such concentration gradients are, however, subject to
unavoidable noise arising from intrinsic biochemical fluctuations. We
therefore investigate how precisely a noisy concentration gradient can
specify positional information.  We analyse noisy one and two gradient
models, where we find that time-averaging of concentration
measurements potentially allows great precision to be achieved even
with remarkably low protein copy numbers. We also find that a
particular choice of the gradient decay length optimizes positional
precision.  Furthermore, in two-dimensional gradients, such as on a
cell membrane, we find that positional precision is substantially
independent of the size of the concentration measuring apparatus. We
apply our results to a number of examples in both cell and
developmental biology, including cell division positioning in bacteria
and yeast, as well as precision gene expression in {\em Drosophila}.

\section*{Introduction}

To determine position in a biological system, some component within
the system must have a non-uniform spatial distribution. Often this is
achieved through the formation of gradients of protein concentration.
Typically a gradient forms when a protein is manufactured/injected
within a small region, and subsequently spreads and decays \cite{W}.
By measuring the local concentration, position relative to the source
can be determined. In developmental biology, where such gradients are
used to control patterns of gene expression, gradient proteins are
called morphogens.  However, intracellular concentration gradients are
also thought to be important for organisation inside single cells.

For a gradient mechanism to be biologically viable, position
determination must be precise and therefore robust to noise.
Variability from one copy of the system to another (e.g. from cell to
cell or embryo to embryo) will certainly compromise positional
precision.  Production and degradation rates can vary, for example,
due to different copy numbers of transcription factors or
proteases. The physical size of the system will also vary and this may
affect proper positioning. Most previous analyses of morphogen
gradients have focused on robustness to changes in these extrinsic
factors 
[2-4] between different copies of
the system. However, there will also be intrinsic noise affecting the
gradient within a single copy of the system, for example due to the
unavoidably noisy nature of the biochemical reactions involved. This
dissection of the fluctuations into extrinsic or intrinsic mirrors
that introduced into the analysis of stochastic gene expression
[5-7]. However, here intrinsic noise
alters not only the overall protein copy numbers (similar to
\cite{Elowitz}), but crucially also the spatiotemporal protein
distribution. Even if all extrinsic variation could be eliminated,
intrinsic biochemical noise would still lead to a fundamental limit to
the precision of position determination, in a similar way to limits on
the precision of protein concentration measurement \cite{Berg,
Bialek}. In this paper, we therefore address the question of how
precisely a concentration gradient can specify positional information,
and calculate the limits on positional precision for a simple, but
biologically relevant, gradient formation mechanism with first order
reaction kinetics.

Quantitative measurements, for example on the Bicoid-Hunchback system
in {\em Drosophila} \cite{bic}, have shown that remarkable positional
precision can sometimes be obtained. For this reason, understanding
the fundamental limits to the precision of concentration gradients is
clearly an important issue in developmental biology. Our results will
be equally relevant to gradients that form within single cells, where
protein copy numbers of a few thousand 
[11-13]
will lead to large density fluctuations.  The properties of
intracellular protein gradients have been studied by Brown and
Kholodenko \cite{Kholodenko}. Recently a number of these gradients
have been observed experimentally in both prokaryotic and eukaryotic
systems. The bacterial virulence factor IcsA forms a polar gradient on
the cell membrane of {\em Shigella flexneri} \cite{Icsa}. MipZ in {\em
Caulobacter} forms polar gradients to aid division site selection
\cite{MipZ}. In {\em B. subtilis}, the MinCD complex also forms polar
gradients in order to direct division site selection to the mid-plane
of the cell \cite{MinCD,how04}. In {\em E. coli} the oscillatory
dynamics of the Min proteins creates a time-averaged gradient that
directs cell division placement 
[18-24]. 
Using mechanisms of this sort, division site placement in bacteria can
achieve an impressive precision of $\pm1\%$ of the cell length
\cite{Ecoli_div, Bsub_div}. Cell division in eukaryotic cells is also
believed to be regulated by concentration gradients. For example, in
fission yeast, the protein Pom1p forms a cortical concentration
gradient emanating from a cell tip, thereby restricting the cell
division protein Mid1p to the cell centre \cite{Paoletti,Chang}. In
eukaryotic cells, gradients of the Ran and HURP proteins aid the
formation of the mitotic spindle by biasing microtubule growth towards
the chromosomes 
[29-33]. 
Gradients may also play a role in the localization of Cdc42
activation, thereby permitting a coupling between cell shape and
protein activation \cite{Odde,Howarddisp}.

Suppose that a biological system needs to identify a particular
position along its length, such as the mid-plane to ensure symmetrical
cell division. As concrete examples, MipZ and the MinCD complex act by
displacing the essential cell division protein FtsZ from the cell
membrane. Since the concentrations of MipZ/MinCD are higher near the
cell poles, FtsZ accumulates near the cell centre. Below some critical
threshold of MinCD or MipZ concentration, enough FtsZ will presumably
accumulate to form the division apparatus. The locations where the
concentration gradient crosses these thresholds mark positions within
the cell. In our analysis we will simply postulate the existence of
such well-defined critical thresholds, where the gradient sharply
switches a downstream signal from on to off. Clearly any real gradient
cannot act as such a sharp switch -- in reality a certain amount of
smearing is inevitable. Furthermore, there will be additional noise in
the process of actually measuring the concentration due both to the
binding of the gradient proteins to the receptor molecules
\cite{Berg,Bialek}, and also to the downstream reactions that process
this incoming signal
[5-7,36-38]. In
general, the noise of the output signal of a processing network can be
written as the sum of a contribution from the noise in the input
signal plus a contribution from the reactions that constitute the
processing network. We assume here that the detector and the
processing network are ideal and do not add any noise to the gradient
input signal. As a result, our calculated variation constitutes a
lower bound; any real gradient signalling system will inevitably have
a lower precision.

We first consider a system with a single planar morphogen source and
linear degradation, thereby producing an exponentially decaying
average concentration profile.  While this model is very simple, it
remains biologically relevant in both developmental and intracellular
contexts. Gradients of Bicoid in {\em Drosophila} and IcsA in {\em
Shigella} have been quantitatively measured and shown to fit this
exponential decay profile on average to high accuracy \cite{bic,
Icsa}. We then calculate the expected distribution of positions where
a noisy gradient crosses a concentration threshold. With typical
cellular copy numbers of a few thousand proteins, the system will be
unable to identify the correct threshold position from a single
measurement. In order to achieve reliable position determination the
concentration must be averaged over time. We show that by averaging
measurements, a biological system is able to achieve precision in
position determination of a few percent of the system size even with
hundreds of protein copies, a result we verify by computer
simulations.  Furthermore, we find that the precision of position
determination is maximised when a particular choice of the gradient
decay length is made. We also show how the precision depends on the
detector size (i.e. the volume over which the density measurement is
made).  For a two dimensional gradient (e.g. on a membrane), the
precision possible after a certain averaging time depends only very
weakly on the detector size. We relate all these results to
experimental measurements of gradients in {\em Shigella} and fission
yeast.

We also consider the ability of gradients from two poles to identify
the centre of the system, as in the MipZ and Pom1p gradients
discussed above. Related designs have also been proposed for the
control of {\it hunchback} positioning in {\em Drosophila}
\cite{H-tW,MH-R-L,H-W-L}. As before, we find that the precision of the
system can be optimised by a particular choice of the decay
length. However, if the threshold position is set at the system
centre, time-averaging improves precision more slowly than in the
single-source model. For subcellular gradients we find that a few
thousand copies of the gradient proteins may therefore be required for
high precision.  Our results strongly constrain the possible
concentrations of gradient proteins in two gradient systems.

\section*{Results}
\subsection*{Single Gradient Model}

We consider a protein gradient which is used to determine a particular
position along the length of a cylindrical system. The system will
have dimension $d=2$ if the gradient is restricted to the membrane, or
$d=3$ if the gradient is in the cytoplasm. We choose the $x$-axis
along the long axis of the system. Position in the remaining
coordinates is denoted by the vector $\mathbf{y}$. For a membrane
system, periodic boundary conditions are appropriate in the
$y$-direction. Otherwise, zero-flux boundaries are used
throughout. The system length is $L$, and the size of the system in
the remaining directions is taken to be $L_\perp$ (so $L_\perp=2\pi
r$, where $r$ is the system radius, for the $2d$ membrane case). A
source on the $x=0$ plane produces proteins at rate $J$ per unit area,
which then diffuse with diffusion constant $D$, and are degraded
uniformly at rate $\mu$. Neglecting fluctuations, the protein
concentration $\rho(x,\mathbf{y},t)$ will be described by
\begin{equation}
\frac{\partial \rho}{\partial t}=D\nabla^2\rho-\mu\rho+J\delta(x). 
\label{pde2}
\end{equation}
If $L\gg\lambda=\sqrt{D/\mu}$, the characteristic decay length of the
gradient, we find that, at steady state, the density is
\begin{equation}
\rho(x)=\frac{J\lambda}{D}\exp\left(-x/\lambda\right).
\label{av}
\end{equation}
Symmetry dictates that the average density is independent of
$\mathbf{y}$. Gradients with the form (\ref{av}) have been found to
accurately fit quantitatively measured concentration profiles in both
developmental \cite{bic} and subcellular \cite{Icsa} systems.

While we have outlined the model in terms of production and
degradation, (\ref{pde2}) could equally apply to other mechanisms in
which the active protein originates in a single location, but
deactivation occurs uniformly throughout the system. The same equation
would therefore describe a protein which is phosphorylated by a
polar-localised kinase and dephosphorylated by a uniformly distributed
phosphatase, or a protein which is activated by being injected into
the membrane at a pole and deactivated when it dissociates. These
biochemical details do not affect the behaviour of the model.

We suppose that signalling is active where the local gradient protein
concentration is above some threshold value, $\rho_T$, and inactive
otherwise. The average concentration profile for a single gradient,
(\ref{av}), suggests that the system will be divided into a region
$0\leq x<x_T$ where signalling is active, and a region $x_T\leq x\leq
L$ where signalling is not active, with $\rho_T=\rho(x_T)$. However,
noise in the local protein concentration will cause this threshold
position to fluctuate. This noise may come from intrinsic fluctuations
in the diffusion, injection and decay processes, or from extrinsic
factors which produce systematic changes in the boundary position when
comparing one copy of the system to another. Here we consider only
intrinsic biochemical fluctuations.

Protein production and degradation events are considered to be single
molecule reactions with a fixed probability per unit time, and hence
will be Poisson processes.  We also assume that the hopping of
proteins in or out of a particular region of space is governed by
Poisson statistics, thereby generating a diffusive process for
molecular transport. Since the system is linear, the instantaneous
fluctuations in molecular number, $n$, within a volume $(\Delta x)^d$
centred on the position $(x,\mathbf{y)}$ should also obey Poisson
statistics, with
\begin{equation}
\left<n(x)^2\right>-\left<n(x)\right>^2=\left<n(x)\right>.
\end{equation}
In terms of protein density, this becomes
\begin{equation}
\langle(\Delta\rho(x))^2\rangle=\left<\rho(x)^2\right>-
\left<\rho(x)\right>^2
=\frac{\left<\rho(x)\right>}{(\Delta x)^d}. \label{dr1}
\end{equation}
This relation can also be established using more elaborate field
theoretic techniques (see \cite{tauber}). From this expression for the
variation in the density we can compute the width of the threshold
position distribution by expanding about the average threshold
position $x_T$. To leading order, this width is given by
\begin{equation}
w_0=\frac{\Delta\rho(x_T)}{|\left<\rho'(x_T)\right>|}=\sqrt{
\frac{\lambda D}{J(\Delta x)^d}}\exp\left(x_T/2\lambda\right),
\label{w0}
\end{equation}
where $\rho'(x_T)$ denotes the first derivative of the density at
$x=x_T$.

Here we identify $(\Delta x)^d$ as the size of the region in which the
concentration is being measured. For subcellular gradients involved in
positional information, this volume will be determined by the size of
an individual receptor or protein with which the gradient protein
interacts, an example being the interaction between the MinCD and FtsZ
proteins in {\it B. subtilis}. The size of the detector, $\Delta x$,
will then be on a molecular scale. This conclusion still holds even if
the gradient proteins bind cooperatively to the ``detection''
protein/receptor due to the close physical proximity of the bound
molecules. In contrast, however, the cellular length scale will be
much larger, $1\mu$m or bigger. 

Throughout the following analysis we will focus on subcellular
gradients. However, our model can equally be applied to developmental
biology, and we will consider these systems further in the {\em
Discussion}.  As concrete examples we first consider the IcsA polar
gradient on the membrane of the rod-shaped bacterium {\em Shigella}
($L\approx 3\mu$m, $L_\perp\approx 3\mu$m) \cite{Icsa}. IcsA is
exported to the outer membrane at a single pole, after which it
diffuses and undergoes uniform proteolysis by the protease IcsP,
thereby forming an exponential gradient exactly as in our model
\cite{Icsa}. Outer membrane IcsA is then able to recruit actin
nucleation factors. However, a critical concentration of IcsA is
likely needed for actin nucleation: in this way a comet-like actin
tail is generated at only one cell pole thereby generating
unidirectional motility of the pathogen.  A cell will typically have a
few thousand copies of IcsA \cite{Icsa2}, forming a gradient with
$\lambda\approx 0.5\mu$m \cite{Icsa}. We take the detector size to be
$\Delta x=0.01\mu$m, consistent with an interaction between IcsA and
actin nucleation proteins. For diffusion on the cell membrane, we take
$D=1\mu$m$^2s^{-1}$. On the membrane of a cell of this size, there
would be approximately $LL_\perp/(\Delta x)^2\sim10^5$ potential
detector sites, many more than the typical copy number. Even near to
the source pole, detector sites will typically be unoccupied. A
detector region at a distance $x=0.5\mu$m from the highly-occupied
pole will have average occupancy of $\left<n\right>\sim 10^{-1}$. In
the cytoplasm of a similarly sized bacterium, the number of potential
detector sites will be $\sim10^6$, again much larger than the protein
copy numbers typically supported by bacteria.

Similar estimates can be made for single polar gradients in fission
yeast ($L=10\mu$m, $L_\perp=6\mu$m), such as for Pom1p
\cite{Paoletti,Chang}. Here we assume a total of 2000 protein copies
(this concentration has not yet been measured but this number is
plausible \cite{Chang}). We also take $D=1\mu m^2 s^{-1}$ and a decay
length of $\lambda=2\mu$m, parameters that are approximately
consistent with the Pom1p gradient imaged by Padte et al
\cite{Chang}. We again assume that $\Delta x=0.01\mu$m corresponding
to a molecular sized detector, as would be the case if the gradient
protein interacted with other membrane proteins (such as Mid1p)
\cite{Paoletti,Chang}. The typical occupancy of a $\Delta x=0.01\mu$m
site is then $\left<n\right>\sim 10^{-2}$ at $x=2\mu$m from the
source.

As we have seen for both fission yeast and {\em Shigella}, average
detector site occupancies that are very much less than one ensure that
the threshold concentration must necessarily be less than one protein
per site. Since most regions will be devoid of any copies of the
protein, a single instantaneous measurement of the protein density
cannot give a good estimate of the local average
concentration. Additionally, multiple positions where the
concentration crosses $\rho_T$ will be observed simultaneously in such
a measurement since the concentration will be above the threshold
everywhere there is a protein molecule present, and below the
threshold where there is no protein molecule. In order to reliably
determine the average concentration profile the system must therefore
integrate the measured concentration over time.

The noisy concentration profile provided by the gradient protein forms
the input signal that is then time-averaged by a downstream signal
processing network. In general, the mechanism for time averaging is
provided by the lifetimes of the states in the processing network. For
instance, in the case of gene expression, fluctuations in the
occupancy of the promoter by a gene regulatory protein can be filtered
by the lifetime of the mRNA transcript, provided that lifetime is much
longer than the timescale of fluctuations in the promoter occupancy
\cite{Paulsson,Bialek}. Similarly, for subcellular gradients, as in
{\em Shigella}, fluctuations in the gradient can be filtered by the
lifetime of activated receptors/detector proteins or their downstream
products. Provided this time scale is much longer than the
sub-millisecond timescale of the gradient fluctuations, then good
time-averaging can be achieved. Importantly, the reactions in the
downstream network not only time-average the noise of the input
signal, but also add further noise to the signal
[5-7,36-38]. Here,
we focus exclusively on noise in the concentration gradient and do not
model the downstream reactions explicitly, but simply assume they are
noiseless and model them with an effective averaging time.  In
essence we assume that the detector and the network that the process
the gradient signal are ideal and do not add further noise, and are
thus able to time-average the gradient signal in the best possible
way. Our results thus provide a lower bound to the output noise set by
the Poissonian fluctuations of the signalling molecules.

We suppose that averaging over a time-interval $\tau$ we can take
$N_\tau=\tau/\tau_{ind}$ independent measurements of the
concentration. In our ideal case, we then expect that the fluctuations
in the concentration will decrease according to
$1/\sqrt{N_{\tau}}$. Since the width varies linearly with $\Delta\rho$
according to (\ref{w0}), the width will also decrease as
\begin{equation}
w(\tau)\sim w_0\sqrt{\frac{\tau_{ind}}{\tau}}.
\end{equation}
The time-scale $\tau_{ind}$ on which independent measurements can be
made is set in our ideal case solely by the reaction-diffusion
dynamics of the gradient proteins, as discussed in the {\em
Appendix}. For cellular parameter values, the typical reaction
timescale, $1/\mu$, will be much longer than the typical timescale for
diffusion between detector sites, $(\Delta x)^2/D$. Assuming a
molecular sized detector, this latter timescale will be of order
$10^{-4}s$, whereas effective protein lifetimes will typically be
seconds or longer. The Damkohler number for the system, the ratio of
the diffusive and reaction timescales, will therefore be $Da\sim
(\Delta x)^2/\lambda^2\sim10^{-4}$. Since $Da\ll1$, the averaging
time-scale is dominated by diffusive motion.  In $d=3$ we find
$\tau_{ind}\sim(\Delta x)^2/D$. However, in $d=2$, density
correlations decay away more slowly, leading to the appearance of
logarithmic corrections that are weakly dependent on the parameters
$\lambda$ and $\Delta x$. For long averaging times, $\tau\gg1/\mu$,
the width determined from time-averaged measurements will be
\begin{equation}
w(\tau)=k_{2d}\left[\frac{\lambda}{\tau
J}\exp\left(x_T/\lambda\right)
\left(\ln\left(\frac{\lambda^2}{(\Delta x)^2}\right)+\alpha\right)
\right]^{1/2}
\label{wt_2d}
\end{equation}
in $d=2$, and for $d=3$
\begin{equation}
w(\tau)=k_{3d}\left[\frac{\lambda}{\tau
J(\Delta x)}\exp\left(x_T/\lambda\right)\right]^{1/2},
\label{wt_3d}
\end{equation}
where $k_{2d}$, $k_{3d}$ and $\alpha$ are constants.

As we have discussed above, $\Delta x$ will be set by the
concentration detection mechanism. However, in a subcellular context,
$\Delta x$ also sets the highest possible resolution of the
system. Once $w\approx\Delta x$ the cell cannot resolve the target
position with any higher precision. Equation (\ref{wt_2d}) suggests
that in $d=2$, precision dependends only very weakly on the detector
size, through the logarithmic correction factor. Reducing the detector
size will increase the number of independent measurements made in a
given averaging time.  However, since fewer proteins will be measured
by each detector over one averaging period, reducing $\Delta x$ will
therefore increase the instantaneous density fluctuations.  In $d=2$
these two effects will largely cancel. Hence, even if we have
over/underestimated the detector volume, this will have little effect
on the precision of two dimensional gradients, such as IcsA in {\em
Shigella} or Pom1p in fission yeast. In three dimensions, however, $w$
varies as $(\Delta x)^{-1/2}$. Since increasing $\Delta x$ reduces $w$
in both $d=2$ and $d=3$, an optimal strategy would be to choose
$\Delta x$ to match the desired precision in order to minimise the
required averaging time.

Intriguingly, from equations (\ref{wt_2d}) and (\ref{wt_3d}) we find
that there exists an optimal decay length such that precision is
maximised. This result can be understood as follows: for fixed $x_T$,
and for $\lambda\gg x_T$, the value of the $|\left<\rho'(x_T)\right>|$
tends to a constant $J/D$, independent of $x_T$. However, as $\lambda$
increases, $\langle\rho(x_T)\rangle$ increases and therefore the
absolute size of the fluctuations in the density also
increases. Therefore, for large and increasing values of $\lambda$,
$w\propto\langle\sqrt{\rho(x_T)}\rangle/\langle \rho'(x_T)\rangle$
must be increasing. Now if $\lambda$ is small ($\lambda\ll x_T$) and
decreasing, when computing the width
$\propto\langle\sqrt{\rho(x_T)}\rangle/\langle \rho'(x_T)\rangle$ the
presence of the square root means that the numerator decreases much
more slowly than the denominator. Hence the width must again increase
as $\lambda$ is decreased for small $\lambda$. Combining these results
for small and large $\lambda$, the width must have a minimum, optimum
value as a function of $\lambda$. This occurs at $\lambda=x_T$ in
$d=3$. In $d=2$, the optimal decay length is given approximately by
\begin{equation}
\lambda\approx x_T\left(1-\frac{1}{\ln(x_T/(\Delta x))}\right),
\label{w_min}
\end{equation}
where we have retained the first order logarithmic correction. 

In order to examine the biological impact of equation (\ref{wt_2d}) we
again consider the Pom1p membrane gradient in fission yeast
\cite{Paoletti,Chang}, using the parameters described
earlier. Simulations of this example system were performed as
described in the {\em Methods} with on average 100 proteins in the
system. Figures \ref{sgl}A and B show how the measured threshold
position, $\bar{x}$, and width, $w$, vary with averaging time. For
long averaging times the simulation data gives excellent agreement
with (\ref{wt_2d}), with the constants $k_{2d}=0.40\pm 0.02$ and
$\alpha=2.5\pm 0.8$.  Figure \ref{sgl}C shows the $w\sim\tau^{-1/2}$
behaviour predicted in (\ref{wt_2d}), and figure \ref{sgl}D confirms
that the width has a minimum as a function of $\lambda$. The
simulation results are consistent with the position of the minimum
predicted by (\ref{w_min}). Figure \ref{sgl}E shows that the
distribution of measured threshold positions is Gaussian to a good
approximation.

Since the averaging timescale $\tau_{ind}$ in a subcellular system is
of order $\sim10^{-4}s$, time-averaging over a period of minutes can
achieve great precision even with very few copies of the gradient
protein. With the parameter values given above, equation (\ref{wt_2d})
predicts that the position $x_T=2\mu$m can be located to within $\pm
0.5\mu$m within an averaging time $\tau=60s$ even if the system
contains on average only about 20 copies of the protein. $\pm 0.1\mu$m
precision can be achieved in the same averaging time with around 400
copies of the protein, a remarkably high level of precision for such a
low concentration. {\em In vivo} Pom1p gradients may be formed by a
few thousand protein copies, allowing for even greater precision.

However, we can see in figure \ref{sgl}B that for averaging times of
less than about a second, the simulation results are not consistent
with (\ref{wt_2d}). In this regime both $w$ and $\bar{x}$ are equal to
$\lambda$. As discussed above, at very short averaging times the
presence of a particle at any position will cause the time-averaged
concentration to be above $\rho_T$ at that point and hence generally
will generate a threshold crossing. The probability distribution of
threshold measurements, $p(x)$, will therefore follow the probability
distribution of particles. Assuming $L\gg \lambda$ we have
\begin{equation}
p(x)dx=\lambda^{-1} \exp(-x/\lambda) dx.
\label{px}
\end{equation}
The cell will on average estimate the threshold position to be
\begin{equation}
\bar{x}=\int_0^Lxp(x)dx\approx\lambda,
\end{equation}
and measurements will be distributed about this position with variance
\begin{equation}
w^2=\int_0^L(x-\bar{x})^2p(x)dx\approx\lambda^2.
\end{equation}
The system is therefore unable to resolve the correct threshold
position at these short time scales if this is different from
$\lambda$.

Associated with the average concentration at the threshold is a length
scale, $l\sim\rho_T^{-1/d}$, the typical distance between proteins at
this position. The average time for a protein to diffuse this distance
will scale as $l^2/D$. In two dimensions, this time is given by
\begin{equation}
\tau_\times\sim(\left<\rho(x_T)\right>D)^{-1} = (J\lambda)^{-1}
\exp(x_T/\lambda). \label{tx}
\end{equation}
Since $\tau_\times$ is the timescale on which a diffusing particle
first arrives at $x_T$, if $\tau\ll \tau_\times$ there will generally
be no particles detected at $x_T$ in the averaging period. The system
therefore cannot reliably estimate the mean concentration at $x_T$, and
hence cannot precisely identify the threshold position. For
averaging times much greater than $\tau_\times$, on average at least
one particle will be detected at $x_T$. The time-averaged
concentration profile will then approach (\ref{av}), and $\bar{x}$
will approach $x_T$. Hence $\tau_\times$ determines the cross-over
time between the two observed regimes of constant $w$ and $w\propto
\tau^{-1/2}$. Figure \ref{sgl}F shows that the scaling in equation
(\ref{tx}) is also reproduced in our simulations. For the parameter
values above, $\tau_\times=0.3s$, and for a more realistic copy number
of 1000, $\tau_\times=0.03s$.  These timescales are extremely short
compared to cell cycle timescales, but do nevertheless show that some
sort of time averaging is probably essential: a single instantaneous
measurement is unlikely to provide precise positional information. In
fact, as we have seen, averaging over much longer times (tens of
seconds) may be necessary if very high ($1\%$) precision is required.

Simulations of the model in three dimensions were also performed (data
not shown). Similar behaviour was observed in this case, and equation
(\ref{wt_3d}) gave good agreement with the observed width at long
averaging times.

\subsection*{Oppositely directed gradients}

In order to reliably locate the centre of a system, the mechanism
responsible must incorporate information about the overall system size
so that the identified position can scale correctly. A single gradient
characterised by a fixed decay length cannot achieve this. We
therefore examine a system where protein gradients are produced by
sources at both ends, and where the central position is identified as
a concentration minimum.

We modify our earlier model by adding an additional planar source at
$x=L$. This addition is appropriate for modelling cell division
inhibitors, such as MipZ in {\em Caulobacter}, that are injected into
the membrane near both cell poles. However, our model would apply
equally if the two sources are of different repressor proteins (as may
be the case in fission yeast \cite{Paoletti,Chang}), although we do
assume that $J$, $D$ and $\mu$ are the same for both gradients. In
this scenario, signalling activity will be determined by the total
concentration. Without fluctuations, this will be described by
\begin{equation}
\frac{\partial \rho}{\partial t}=D\nabla^2\rho
-\mu\rho+J\delta(x)+J\delta(x-L). \label{pde}
\end{equation}
The steady-state solution is now
\begin{equation}
\rho(x)=\frac{J\lambda}{D}\frac{\cosh((x-L/2)/
\lambda)}{\sinh(L/2\lambda)}, \label{r_dbl}
\end{equation}
which has the expected minimum at $x=L/2$. 

We then suppose that the cell compares the concentration to a
threshold value corresponding to the minimum of the average profile,
$\rho_{min}=\rho(L/2)=\rho_T$. Positions where the concentration is at
or below the threshold are identified as being at the centre of the
cell. While the average steady-state density profile would never
extend below $\rho_{min}$, fluctuations ensure that the concentration
in the region around the centre spends a significant amount of time at
or below the threshold. Around point(s) where
$\langle\rho(x)\rangle=\rho_T$, noise in the protein concentration
will lead to a distribution of threshold crossing positions. We
consider an expansion of the density fluctuations about $x_T=L/2$,
giving, to leading order
\begin{equation}
\Delta\rho(x_T)=\frac{1}{2}\left|
\left<\rho''(x_T)\right>\right|w^2,
\label{drx}
\end{equation}
since any first order term proportional to $\langle\rho'\rangle$
vanishes at $x_T=L/2$. The width is therefore given by
\begin{equation}
w^2=\frac{2\Delta \rho(L/2)}{\left<\rho''(L/2)\right>}.
\label{wsum2}
\end{equation}
Substituting in (\ref{r_dbl}) gives
\begin{equation}
w_0=\left(\frac{4D\lambda^3\sinh(L/2\lambda)}{J(\Delta
x)^{d}}\right)^{1/4}.
\end{equation}

As in the single gradient model, the typical occupancy of the
threshold region will be much less than one. For example, if we take
the parameter values considered previously for the Pom1p gradient in
fission yeast, with 2000 protein copies, the average occupancy of a
detector site at $x=L/2$ will be $\left<n(L/2)\right>\sim 10^{-3}$. We
assume here that Pom1p forms a gradient from both poles. In fact it
may only form a single gradient with another hitherto unidentified
protein forming the second polar gradient
\cite{Paoletti,Chang}. However, as discussed earlier, this detail does
not affect our calculations. As a second example, MipZ in {\em
Caulobacter} ($L=2.5\mu$m, $L_\perp=2\mu$m) is typically present at
about $1000$ copies, and forms two polar gradients with a decay length
$\lambda\approx 0.25\mu$m \cite{MipZ}. The average occupancy at the
centre of this system would be approximately $\left<n(L/2)\right>\sim
10^{-3}$. Averaging measurements of the concentration over time is
therefore required in both cases to obtain precise positional
information. Since the width now goes as
$\left(\Delta\rho\right)^{1/2}$, as shown in (\ref{wsum2}), we expect
\begin{eqnarray}
w(\tau)&=&w_0\left(\frac{\tau_{ind}}{\tau}\right)^{1/4}\nonumber \\
&=&\begin{cases}
\tilde{k}_{2d}\left[\frac{\lambda^3}{\tau
J}
\sinh(L/2\lambda)\left(\ln\left(\frac{\lambda^2}{(\Delta
x)^2}\right)+\tilde{\alpha}\right)\right]^{1/4}& \ \text{in $d=2$} \\
\tilde{k}_{3d}\left[\frac{\lambda^3}{\tau
J(\Delta x)}\sinh(L/2\lambda)\right]^{1/4}& \ \text{in $d=3$}
\end{cases},\label{dbl_wt}
\end{eqnarray}
where $\tilde{k}_{2d}$, $\tilde{\alpha}$ and $\tilde{k}_{3d}$ are
constants.  Averaging proceeds much more slowly than previously, with
a $\tau^{-1/4}$ dependence. This follows directly from the vanishing
of the first derivative at the average threshold position. In $d=3$,
and for $\lambda\ll L$, equation (\ref{dbl_wt}) predicts that $w$ will
be minimised when $\lambda\approx L/6$ is chosen. In $d=2$ logarithmic
corrections again alter this result slightly, with the optimal decay
length now occurring at
\begin{equation}
\lambda\approx\frac{L}{6}\left(1-\frac{1}{3\ln(L/6(\Delta
x))}\right),
\end{equation}
where we have included the leading logarithmic correction.  This
result arises for similar reasons as in the single gradient
model. For the Pom1p gradient imaged by Padte et al \cite{Chang}, the
decay length is observed to be $1-1.5\mu$m, comparable to this optimal
decay length of about $1.5\mu$m for a $10\mu$m cell.

\begin{figure}
\includegraphics[width=1.0\textwidth]{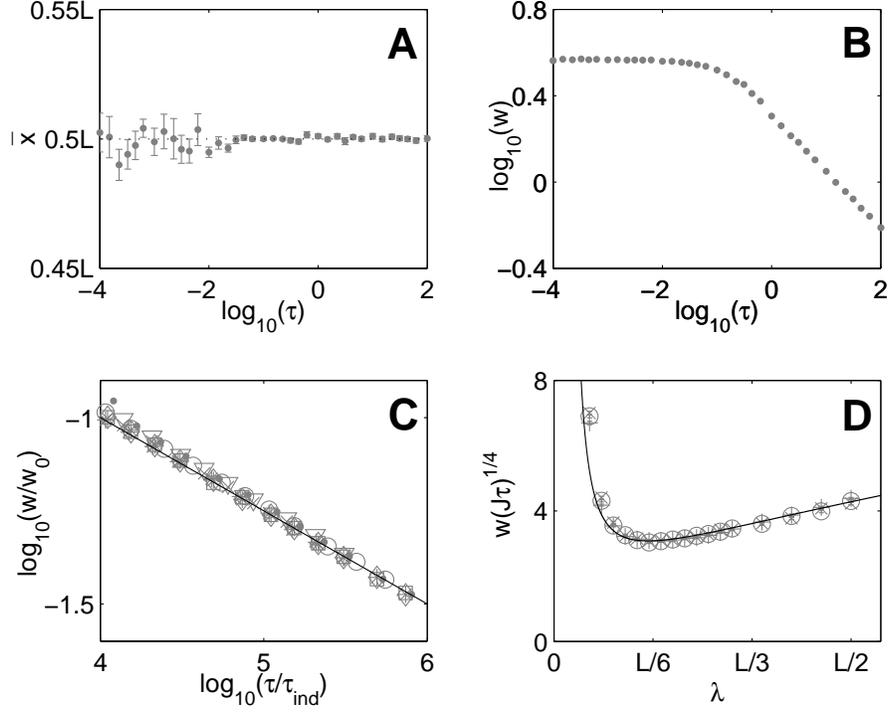}
\caption{Two gradient model in $2d$. A. The mean threshold position
fluctuates about $L/2$ due to the symmetry of the system.
B. Variation of the width $w$ as a function of averaging time. C. Data
collapse of the width as a function of averaging time, at long times,
for a range of parameter values. The full line shows (\ref{dbl_wt})
with $\tilde{k}_{2d}=0.63$ and $\tilde{\alpha}=2.5$. $*$ indicates the
standard parameter values. For the other data sets parameter values
were changed as follows: $\circ$: $D=0.5\mu$m$^2s^{-1}$; $\square$:
$J=9\mu$m$^{-1}s^{-1}$; $\times$: $\Delta x=0.02\mu$m; $\bullet$:
$\mu=1s^{-1}$; $+$: $\mu=0.25s^{-1}$; $\diamond$: $L=7.5\mu$m;
$\triangledown$: $L=15\mu$m and $\Delta x=0.02\mu$m. D. Plot of width
as a function of decay length for averaging times $\times$:
$\tau=30s$; $\circ$: $\tau=45s$; and $+$: $\tau=60s$. The full line
shows the prediction from equation (\ref{dbl_wt}).}
\label{dbl}
\end{figure}

We simulated our model in two dimensions with representative parameter
values for fission yeast membrane gradients. We used $\mu=0.36s^{-1}$
chosen to give $\lambda=1.67\mu$m, and $J=6\mu$m$^{-1}s^{-1}$ giving
on average 200 protein copies in total. Figure \ref{dbl} shows the
results of these simulations. Again we observe two distinct
regimes. At averaging times longer than about a second, there is
excellent agreement with equation (\ref{dbl_wt}), as we can see in
figure \ref{dbl}C. Fitting to the simulation results we find $\tilde{
k}_{2d}=0.63\pm 0.02$ and $\tilde{\alpha}=2.5\pm 1.0$. Figure
\ref{dbl}D confirms the existence of the optimal decay length in our
simulations.

Since the width decays as $\tau^{-1/4}$ for this system, longer
averaging times and/or higher protein copy numbers are required than
in the single gradient model to achieve high precision.  Intrinsic
biochemical noise may therefore strongly constrain systems of this
type. In order for the yeast-membrane gradient considered above to
achieve precision of $\pm 5\%$ of the cell length after averaging for
one minute, about 800 protein copies are required. Therefore, in the
absence of any other positioning mechanisms, the Pom1p gradient will
require $\sim1000$ protein copies or more to precisely direct the
location of cell division. We estimate that the MipZ gradient in {\em
Caulobacter}, with 1000 protein copies, would be able to locate the
cell centre to within $\pm5\%$ of $L$ after approximately $\tau=2s$.
However, since precision only improves as $\tau^{-1/4}$, averaging
over $\tau=20$ minutes would be required for the same system to
achieve $\pm1\%$ accuracy.

\section*{Discussion}

Noise in biochemical processes within a cell will lead to fluctuations
in protein concentration gradients, and hence also to variation in the
position where these gradients cross a particular threshold
value. These fluctuations therefore place a limit on the potential
precision of position determination mechanisms relying on
concentration gradients alone. In subcellular systems with protein
copy numbers in the thousands, this noise will be sufficiently large
that position cannot be determined reliably from a single measurement
of the density profile. In order to determine position to within a few
percent, a precision achieved by some subcellular systems, the protein
concentration must be averaged over time. For a single subcellular
membrane gradient, we have seen that by averaging over a period of a
minute, excellent precision can potentially be achieved with only a
few hundred protein copies. This remarkable precision is due to the
sub-millisecond diffusive time-scale on which time-averaging occurs.
Precise identification of the cell mid-plane by gradients emanating
from both poles requires longer averaging times or higher copy
numbers, since larger fluctuations result from the vanishing first
derivative of the average concentration at the system
centre. Intrinsic biochemical noise may therefore be a strong
constraint on subcellular two-gradient positioning systems, dictating
that the copy numbers be sufficiently high to suppress fluctuations.

So far we have focused almost exclusively on fluctuations in
subcellular gradients, however our results are also applicable to
developmental biology and we wish to briefly comment on this
application. Here the appropriate length scales are usually much
longer, on the order of hundreds of micrometers in {\em
Drosophila}. Moreover, the gradients affect patterns of gene
expression through the binding of gradient molecules to DNA regulatory
sequences inside individual nuclei. For example, in {\em Drosophila},
where exponential gradients have been quantitatively measured for
Bicoid \cite{bic}, Bicoid binds cooperatively to {\em hunchback}
regulatory DNA. In this case we again expect molecular-scale effective
measuring volumes, with $\Delta x\sim 0.01\mu$m being a reasonable
order of magnitude.  We next assume purely Poisson statistics for the
fluctuations: this is a stronger assumption than for our earlier
subcellular gradients, as there will be additional complications
arising, for example, from the import/export of morphogens from
nuclear compartments. However, if diffusive noise is dominant then
Poisson statistics will be retained and we can expect our earlier
analysis to apply, although with one important distinction. Instead of
$\Delta x$ setting the maximal possible precision, this will now be
set by the size of individual nuclei (prior to cellularization), since
we expect relatively homogeneous gene expression within a single
nuclear volume. A single nucleus in {\it Drosophila} has a length
scale of around $10\mu$m, still much smaller than the decay length of
the gradient of $\lambda\sim 100\mu$m, allowing for high precision
gene expression \cite{bic}. Using the {\it Drosophila} Bicoid gradient
as an example, we use $L=500\mu$m, $L_{\perp}=100\mu$m, and estimate
$D=10\mu$m$^2s^{-1}$ and $\mu=10^{-3}s^{-1}$, giving
$\lambda=100\mu$m, consistent with experiment \cite{bic}.  Assuming a
high copy number of $10^7$ per embryo (we are not aware of
experimental constraints on this figure), gives $J\sim
1\mu$m$^{-2}s^{-1}$. For a single gradient in three dimensions, we
find that about a 5 minute averaging time is required to bring the
error down to plus or minus a single nuclear length.  For a two
gradient model in three dimensions, longer averaging times on the
order of an hour are required to reduce the centre-finding positional
error to plus or minus about 2 nuclear lengths. Since gene expression
may need to be controlled on shorter timescales than this, other
designs, for example using {\it interacting} gradients
\cite{H-tW,MH-R-L}, may be required for high precision centre finding
(see also below). The effects of the optimum gradient length scale
will also be interesting to probe in a developmental biology
context. However, our simple analysis may be complicated by the
multiple roles played by many morphogens: for example, Bicoid not only
activates {\it hunchback}, but it also helps to regulate pair-rule
genes, such as Even-skipped. Nevertheless, it is interesting to note
that the Bicoid gradient length scale $\lambda\sim 100 \mu$m
\cite{bic} is not too far away from the $L/6$ optimum for a two
gradient case, and in a single gradient context will offer maximal
precision well into the anterior half of the embryo.

Up to this point we have only considered systems with first order
degradation. Morphogen gradients with nonlinear decay have also been
proposed \cite{Barkai_nlin}. This nonlinearity will lead to
non-Poissonian density fluctuations, which may significantly change
the observed behaviour. England and Cardy \cite{E-C} have previously
calculated the response of a gradient with nonlinear decay to one
source of biochemical noise, namely a fluctuating production
rate. However, they calculated the change to the average gradient,
while fluctuations about this average may also be important. It would
certainly be of interest to compare the performance of linear and
nonlinear degradation mechanisms in more detail. Centre-finding
mechanisms with interactions have also been proposed \cite{H-tW,
MH-R-L}. In these models position is determined from the combined
gradient of two proteins, which will be steep around the system centre
due to an interaction between the two gradients. These mechanisms may
therefore be able to achieve greater precision for mid-point
determination than the noninteracting mechanism considered here.

Throughout this work we have assumed that the gradient protein
concentration fluctuates about a steady-state profile, and hence
averaging over a longer time will give a more precise estimate of the
average profile.  For a subcellular system, the steady-state gradient
will develop over timescales of less than about a minute, due to the
micrometer length scales involved. This timescale is short compared to
the cell cycle time, which ranges from tens of minutes up to many
hours. For this reason we expect that subcellular gradients will be in
steady-state and therefore that our analysis will be directly
applicable.  However, in developmental biology, the effective
lifetimes will likely be much longer, and the gradient may take hours
to fully reach steady-state. Moreover, a number of developmental
biology systems are known to respond to a morphogen gradient that has
not reached steady-state 
[42-44]. A further complication
is the possibility of gradient formation by non-Fickian diffusion
\cite{Barkai_diff}, where there is no steady-state at all.  The model
considered in this paper does not take into account time-varying
average gradients. If the average gradient is evolving, a longer
averaging period will not necessarily lead to improved
precision. Clearly, more work will be required to understand how such
dynamically evolving systems are able to yield precise positional
information and filter out fluctuations. Nevertheless, we do note that
two gradient systems of the kind analyzed here are naturally able to
locate the system centre even without being in steady-state, due to
the symmetry of the system \cite{H-tW}. The positional variations in
such a non-steady-state scenario will not be the same as calculated
here, but our analysis does form a first step towards the analysis of
these more complex systems.

\section*{Methods}
\label{meth}

\subsection*{Calculation of $\tau_{ind}$}

We have assumed in our analysis that during the time-averaging process
we are taking independent measurements at intervals of
$\tau_{ind}$. However, in both real biological systems and in our
simulations, measurements can generally be taken at much shorter
intervals than this, leading to correlations between consecutive
measurements. For a series of correlated measurements taken at time
intervals $\delta t$ over a period $0\leq t\leq\tau$, with
$\tau\gg\delta t$, the expected error for the time-averaged
concentration at position $\mathbf{x}$,
$(\Delta\rho(\mathbf{x},\tau))^2$, is given by \cite{M-Binder}
\begin{equation}
(\Delta\rho(\mathbf{x},\tau))^2=\frac{\delta
t}{\tau}(\Delta\rho(\mathbf{x},0))^2\left[
1+\frac{2}{\delta t}\int_{0}^{\tau}\left(1-\frac{t}{\tau}\right)C(t)dt
\right],
\end{equation}
where $(\Delta\rho(\mathbf{x},0))^2$ is the variance of a single
measurement,
\begin{equation}
(\Delta\rho(\mathbf{x},0))^2=\left<\rho(\mathbf{x},
0)^2\right>-\left<\rho(\mathbf{x},0)\right>^2,
\end{equation}
and $C(t)$ is the normalized density correlation function,
\begin{equation}
C(t)=\frac{\left<\rho(\mathbf{x},t)
\rho(\mathbf{x},0)\right>-\left<\rho(\mathbf{
x},0)\right>^2}{\left<\rho
(\mathbf{x},0)^2\right> - \left<\rho(\mathbf{x},0)\right>^2}. \label{C_def}
\end{equation}
We therefore define the timescale $\tau_{ind}$ to be 
\begin{equation}
\tau_{ind}(\tau)=2\int_{0}^{\tau}\left(1-\frac{t}{\tau}\right)C(t)dt,
\label{tind}
\end{equation}
and assuming $\tau_{ind}\gg\delta t$ we recover
\begin{equation}
\Delta\rho(\mathbf{x},\tau)=\Delta\rho(\mathbf{x},0)
\left(\frac{\tau_{ind}(\tau)
}{\tau}\right)^
{1/2}.
\end{equation}
For $N$ independent measurements of the density, we would expect the
error to decline as $N^{-1/2}$. For large enough values of
$\tau_{ind}(\tau)$, where $\tau_{ind}$ becomes independent of $\tau$,
we can therefore interpret $\tau_{ind}$ as the time-interval required
for successive measurements to be independent.

The next step of the calculation is to compute the correlation
function $C(t)$ appropriate for our model. For pure diffusion, we
expect:
\begin{eqnarray}
C(t)\sim 1 \qquad & {\rm for} & t \ll \frac{(\Delta x)^2}{D} \\
C(t)\sim \left(\frac{(\Delta x)^2}{Dt}\right)^{d/2} 
\qquad & {\rm for} & t \gg \frac{(\Delta x)^2}{D}. 
\end{eqnarray}
On time scales $t\ll (\Delta x)^2/D$ the system remains perfectly
correlated as there has been insufficient time for particles to hop
away to neighboring sites. However, for $t\gg (\Delta x)^2/D$, an
algebraically decaying correlation function is found, characteristic
of diffusion. However, we also need to incorporate the effects of
spontaneous decay that occur independently of the diffusive
motion. Adding decay to the system simply alters the correlation
functions by a multiplicative factor of $\exp(-\mu t)$. We now
substitute this full form into the definition of $\tau_{ind}$
(\ref{tind}). In the biologically relevant limits where $\tau\gg
(\Delta x)^2/D$ and $1/\mu \gg (\Delta x)^2/D$, we find, for $d=2$
\begin{eqnarray}
\tau_{ind}\sim\frac{(\Delta x)^2}{D}\left(\ln\left(\frac{
D\tau}{(\Delta x)^2}\right)+{\rm constant}\right) \label{t_short}
\end{eqnarray}
for $\mu\tau\ll1$, and 
\begin{eqnarray}
\tau_{ind}\sim\frac{(\Delta x)^2}{D}\left(\ln\left(\frac{\lambda^2
}{(\Delta x)^2}\right)+{\rm constant}\right) \label{t_long}
\end{eqnarray}
for $\mu\tau\gg1$. In a three-dimensional system we find
\begin{equation}
\tau_{ind}\sim\frac{(\Delta x)^2}{D}.
\end{equation}

For the parameter values considered in our simulations we do not
observe the logarithmic $\tau$-dependence in the width predicted by
(\ref{t_short}).  In the single gradient simulations this is because,
at short times $\tau\ll\tau_\times$, we enter the constant $w\sim
\lambda$ regime. For the parameter values used, the transition from
$w\sim\lambda$ at $\tau\ll\tau_\times\approx0.3s$ to the long time
behaviour (\ref{wt_2d}) for $\tau\gg 1/\mu\approx4s$ overwhelms the
small logarithmic effect. If the production rate $J$ were increased
significantly, $\tau_\times\propto J^{-1}$ would be reduced and the
$\ln \tau$ regime would become accessible since the $\tau_\times$ and
$1/\mu$ timescales would then become better separated. However, even
in this case, the logarithmic variation in (\ref{t_short}) is
intrinsically weak, and will likely have a negligible effect in a
biological context.

\subsection*{Simulations}

Stochastic simulations were performed on a two-dimensional square
lattice with $N_x=L/\delta x$ sites in the $x$-direction and
$N_y=L_\perp/\delta x$ sites in the $y$-direction, where $\delta
x=0.01\mu$m is the lattice spacing. The detector size $\Delta x$ was
normally set equal to $\delta x$ except for cases where the detector
size was varied, in which case $\Delta x$ was set to be a multiple of
$\delta x$. Zero-flux boundaries were implemented at $x=0$ and $x=L$,
and a periodic boundary was used to connect $y=0$ with $y=L_\perp$.  A
fixed time step, $\delta t=2.5\times10^{-5}s$, was chosen so that for
the given diffusion constant the total probability of diffusion out of
a site in all directions approached 1. However, a timestep $5$ times
smaller was also tested with no effect on any of the results. For each
$x=0$ site, particles were injected at each time step in a Poisson
process with mean $j=J\delta x\delta t$. In the two-gradient model,
particles were also added at $x=L$ in an identical but uncorrelated
process. Diffusion and decay were also treated as Poisson processes,
with hopping and decay probabilities of $D \delta t/(\delta x)^2$ and
$\mu\delta t$ per particle respectively. Simulations were initialised
with the mean number of particles in the system, $JL_\perp/\mu$ for
the one-gradient model or twice this value for the two-gradient model,
with a probability distribution that followed the average density
distribution.

The mean occupancy for each detector site was calculated over the
averaging period, $\tau$. For each site this mean occupancy was
compared with each neighbouring site. If one occupancy was above the
threshold and the other below, this boundary was identified as a
threshold crossing position. This process was repeated for many
averaging periods, ranging from $10^5$ repeats for short averaging
times to 500 repeats for very long averaging times, to generate a
distribution of crossing positions throughout the system.  Threshold
crossings in both the $x$- and $y-$directions were observed. We found
that the distributions as a function of $x-$position of these two
types of crossing were the same.  For each row of sites, $x=0$ to
$x=L$ at fixed $y$, the mean (``measured threshold'') and
root-mean-squared deviation (``width'') of the threshold distribution
from many averaging periods were calculated independently. In the
figures we plot the mean of these two quantities across the different
$y$-values within the system, with error bars of one standard
deviation.

For the single-source model the standard parameter values used in the
simulations were as follows: $L=10\mu$m, $L_\perp=6\mu$m, $D=1\mu$m$
^2s^{-1}$, $\mu=0.25s^{-1}$, $J=4.17\mu$m$^{-1}s^{-1}$, $\Delta
x=0.01\mu$m, $x_T=2\mu$m. To generate the data collapse in figures
\ref{sgl}C and F, simulations were also performed with: $D=0.5\mu
m^2s^{-1}$; $J=6.25\mu$m$^{-1}s^{-1}$; $\Delta x=0.02\mu$m;
$\mu=1s^{-1}$; $\mu=0.11s^{-1}$; $x_T=1\mu$m; $x_T=3\mu$m. For the
two-source model, standard parameters were the same as above except
$\mu=0.36s^{-1}$ and $J=6\mu$m$^{-1}s^{-1}$. In figure \ref{dbl}C data
are also shown with: $D=0.5\mu$m$^2s^{-1}$; $\mu=1s^{-1}$;
$\mu=0.25s^{-1}$; $J=9\mu$m$^{-1}s^{-1}$; $\Delta x=0.02\mu$m;
$L=7.5\mu$m; $L=15\mu$m and $\Delta x=0.02\mu$m.

\subsection*{Funding}
F.T. is supported by the EPSRC, and M.H. by The Royal Society.

\newpage
\begin{figure}
\includegraphics[width=1.0\textwidth]{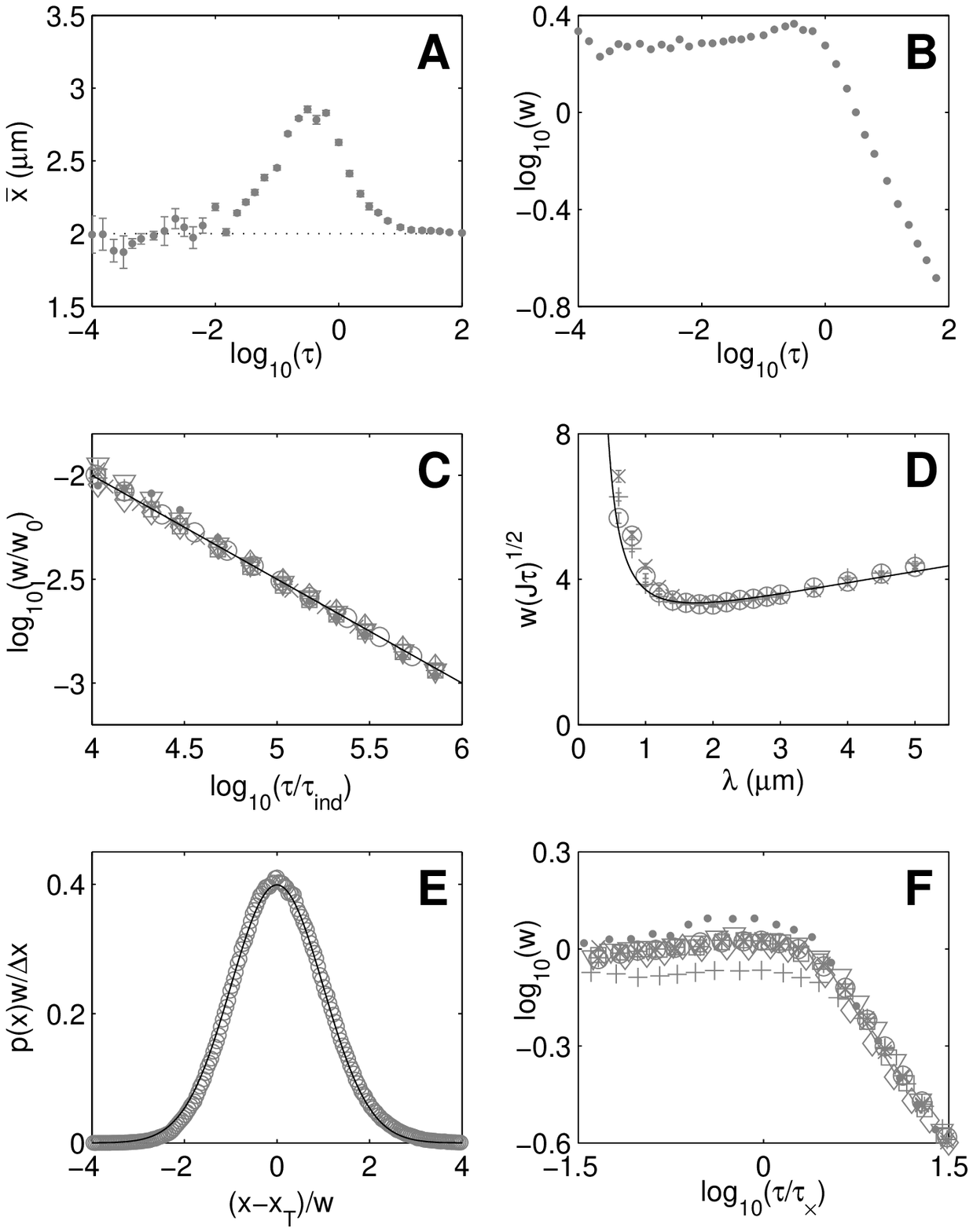}
\end{figure}
\begin{figure}
\caption{Single gradient model in $2d$. A. Variation of the estimated
threshold position with averaging time, with $x_T=2\mu$m and
$\lambda=2\mu$m.  B. Variation of the width as a function of averaging
time. C. Data collapse of the width at large $\tau$ for a range of
parameter values.  Full line shows the prediction of equation
(\ref{wt_2d}) with $k_{2d}=0.40$ and $\alpha=2.5$. D.  $w(\tau)$ as a
function of decay length, with $x_T=2\mu$m. Results for three
different averaging times are shown: $\times$: $\tau=10s$; $\circ$:
$\tau=15s$; and $+$: $\tau=22.5s$. The full line shows the prediction
from equation (\ref{wt_2d}). At large $\lambda$ the simulation results
deviate from the prediction since the assumption that $L\gg\lambda$ is
no longer valid. E. Plot of the probability distribution for measuring
the threshold at position $x$ with an averaging time $\tau=45s$. The
full line shows a normal distribution. F. Scaling of the cross-over
time, $\tau_\times$, according to equation (\ref{tx}). In figures A.,
B. and E. the standard parameter values given in the text were
used. In figures C. and F., $*$ indicates the standard parameter
values. For the other data sets one parameter value was changed as
follows: $\circ$: $D=0.5\mu$m$^2s^{-1}$; $\square$:
$J=6.25\mu$m$^{-1}s^{-1}$; $\times$: $\Delta x=0.02\mu$m; $\bullet$:
$\mu=1s^{-1}$; $+$: $\mu=0.11s^{-1}$; $\diamond$: $x_T=1\mu$m;
$\triangledown$: $x_T=3\mu$m.}
\label{sgl}
\end{figure}

\end{document}